	\documentclass[conference]{IEEEtran}
	\usepackage{multicol}
	\usepackage{lipsum,graphicx}
	\usepackage{float}
	\usepackage{epstopdf}
	\usepackage{ifthen}
	\usepackage[latin1]{inputenc}
	
	\usepackage{amssymb}
	\usepackage[cmex10]{amsmath}
	\usepackage{dcolumn}
	\usepackage{fixltx2e}
	\usepackage[nolist]{acronym}
	
	\DeclareGraphicsExtensions{.pdf,.jpeg,.png,.eps}
	\usepackage{booktabs}
	
	\usepackage[caption=false,font=footnotesize]{subfig}
	\usepackage{comment}
	\usepackage{color}
	\usepackage{tikz}
	\usepackage{setspace}
	\usepackage{mathrsfs}          
	\usepackage{amsmath, amssymb, amsthm}
	\usepackage{amsfonts}
	\usepackage{mathrsfs}
	\usepackage{multirow}        
	\usepackage{framed}
	\usepackage{psfrag}
	\usepackage{graphicx}
	\usepackage{caption}
	\usepackage{algpseudocode} 
	\usepackage{empheq}

	\newcolumntype{d}[1]{D{.}{\cdot}{#1} }

	\usepackage{bm}
\usepackage{mathtools}
\usepackage{siunitx}

\newcommand{\PreserveBackslash}[1]{\let\temp=\\#1\let\\=\temp}
\newcolumntype{C}[1]{>{\PreserveBackslash\centering}p{#1}}
\newcolumntype{R}[1]{>{\PreserveBackslash\raggedleft}p{#1}}
\newcolumntype{L}[1]{>{\PreserveBackslash\raggedright}p{#1}}

\newcommand{\figref}[1]{Fig.~\ref{#1}}
\newcommand{\secref}[1]{Section~\ref{#1}}

\newcommand{\compl}{\mathbb{C}}         

\newcommand{\ma}  [1]{ \bm{#1} } 

\newcommand{\Norm}[1]  { \left\| #1 \right\|  }

\newcommand{\Ex}[1]{\mathrm{E}\left[ #1\right]} 

	\usepackage{lipsum}
	\usepackage{mathtools}
	\usepackage{cuted}
	
\begin{document}
	\begin{acronym}
\acro{1G}{first generation}
\acro{2G}{second generation}
\acro{3G}{third generation}
\acro{3GPP}{Third Generation Partnership Project}
\acro{4G}{fourth generation}
\acro{5G}{fifth generation}
\acro{6G}{sixth generation}
\acro{802.11}{IEEE 802.11 specifications}
\acro{A/D}{analog-to-digital}
\acro{AC}{alternating current }
\acro{ADC}{analog-to-digital}
\acro{AMC}{adaptive modulation and coding}
\acro{AM}{amplitude modulation}
\acro{AP}{access point}
\acro{AR}{augmented reality}
\acro{ASIC}{application-specific integrated circuit}
\acro{ASIP}{Application Specific Integrated Processors}
\acro{AWGN}{additive white Gaussian noise}
\acro{BLER}{block error rate}
\acro{BCJR}{Bahl, Cocke, Jelinek and Raviv}
\acro{BER}{bit error rate}
\acro{BFDM}{bi-orthogonal frequency division multiplexing}
\acro{BPSK}{binary phase shift keying}
\acro{BWP}{bandwidth part}
\acro{BS}{base stations}
\acro{CA}{carrier aggregation}
\acro{CAF}{cyclic autocorrelation function}
\acro{Car-2-x}{car-to-car and car-to-infrastructure communication}
\acro{CAZAC}{constant amplitude zero autocorrelation waveform}
\acro{CB-FMT}{cyclic block filtered multitone}
\acro{CCDF}{complementary cumulative density function}
\acro{CDF}{cumulative density function}
\acro{CDMA}{code-division multiple access}
\acro{CFO}{carrier frequency offset}
\acro{CIR}{channel impulse response}
\acro{CM}{complex multiplication}
\acro{COFDM}{coded-\acs{OFDM}}
\acro{CoMP}{coordinated multi point}
\acro{COQAM}{cyclic OQAM}
\acro{CP}{cyclic prefix}
\acro{CPE}{common phase error}
\acro{CR}{cognitive radio}
\acro{CRC}{cyclic redundancy check}
\acro{CRLB}{Cram\'{e}r-Rao lower bound}
\acro{CS}{cyclic suffix}
\acro{CSI}{channel state information}
\acro{CSMA}{carrier-sense multiple access}
\acro{CWCU}{component-wise conditionally unbiased}
\acro{D/A}{digital-to-analog}
\acro{D2D}{device-to-device}
\acro{DCI}{downlink control information}
\acro{DAC}{digital-to-analog}
\acro{DC}{direct current}
\acro{DFE}{decision feedback equalizer}
\acro{DFT}{discrete Fourier transform}
\acro{DL}{downlink}
\acro{DMT}{discrete multitone}
\acro{DNN}{deep neural network}
\acro{DSA}{dynamic spectrum access}
\acro{DSL}{digital subscriber line}
\acro{DSP}{digital signal processor}
\acro{DTFT}{discrete-time Fourier transform}
\acro{DVB}{digital video broadcasting}
\acro{DVB-T}{terrestrial digital video broadcasting}
\acro{DWMT}{discrete wavelet multi tone}
\acro{DZT}{discrete Zak transform}
\acro{E2E}{end-to-end}
\acro{EE} {energy efficiency}
\acro{eNodeB}{evolved node b base station}
\acro{E-SNR}{effective signal-to-noise ratio}
\acro{EVD}{eigenvalue decomposition}
\acro{FBMC}{filter bank multicarrier}
\acro{FD}{frequency-domain}
\acro{FDD}{frequency-division duplexing}
\acro{FDE}{frequency domain equalization}
\acro{FDM}{frequency division multiplex}
\acro{FDMA}{frequency-division multiple access}
\acro{FEC}{forward error correction}
\acro{FER}{frame error rate}
\acro{FFT}{fast Fourier transform}
\acro{FIR}{finite impulse response}
\acro{FM}		{frequency modulation}
\acro{FMT}{filtered multi tone}
\acro{FO}{frequency offset}
\acro{F-OFDM}{filtered-\acs{OFDM}}
\acro{FPGA}{field programmable gate array}
\acro{FSC}{frequency selective channel}
\acro{FS-OQAM-GFDM}{frequency-shift OQAM-GFDM}
\acro{FT}{Fourier transform}
\acro{FTD}{fractional time delay}
\acro{FTN}{faster-than-Nyquist signaling}
\acro{GFDM}{generalized frequency division multiplexing}
\acro{GFDMA}{generalized frequency division multiple access}
\acro{GMC-CDM}{generalized	multicarrier code-division multiplexing}
\acro{GNSS}{global navigation satellite system}
\acro{GS}{guard symbols}
\acro{GSM}{Groupe Sp\'{e}cial Mobile}
\acro{GUI}{graphical user interface}
\acro{H2H}{human-to-human}
\acro{HW}{hardware}
\acro{H2M}{human-to-machine}
\acro{HTC}{human type communication}
\acro{I}{in-phase}
\acro{ID}{identifier}
\acro{i.i.d.}{independent and identically distributed}
\acro{IB}{in-band}
\acro{IBI}{inter-block interference}
\acro{IC}{interference cancellation}
\acro{ICI}{inter-carrier interference}
\acro{ICT}{information and communication technologies}
\acro{ICV}{information coefficient vector}
\acro{IDFT}{inverse discrete Fourier transform}
\acro{IDMA}{interleave division multiple access}
\acro{IEEE}{institute of electrical and electronics engineers}
\acro{IF}{intermediate frequency}
\acro{IFFT}{inverse fast Fourier transform}
\acro{IoT}{Internet of Things}
\acro{IOTA}{isotropic orthogonal transform algorithm}
\acro{IP}{internet protocole}
\acro{IP-core}{intellectual property core}
\acro{ISDB-T}{terrestrial integrated services digital broadcasting}
\acro{ISDN}{integrated services digital network}
\acro{ISI}{inter-symbol interference}
\acro{ITU}{International Telecommunication Union}
\acro{IUI}{inter-user interference}
\acro{LAN}{local area netwrok}
\acro{LEO}{low Earth orbit}
\acro{LLR}{log-likelihood ratio}
\acro{LMMSE}{linear minimum mean square error}
\acro{LNA}{low noise amplifier}
\acro{LO}{local oscillator}
\acro{LOS}{line-of-sight}
\acro{LoS}{line of sight}
\acro{LP}{low-pass}
\acro{LPF}{low-pass filter}
\acro{LS}{least squares}
\acro{LTE}{long term evolution}
\acro{LTE-A}{LTE-Advanced}
\acro{LTIV}{linear time invariant}
\acro{LTV}{linear time variant}
\acro{LUT}{lookup table}
\acro{M2M}{machine-to-machine}
\acro{MA}{multiple access}
\acro{MAC}{multiple access control}
\acro{MAP}{maximum a posteriori}
\acro{MC}{multicarrier}
\acro{MCA}{multicarrier access}
\acro{MCM}{multicarrier modulation}
\acro{MCS}{modulation coding scheme}
\acro{MF}{matched filter}
\acro{MF-SIC}{matched filter with successive interference cancellation}
\acro{MIMO}{multiple-input, multiple-output}
\acro{MISO}{multiple-input single-output}
\acro{ML}{machien learning}
\acro{MLD}{maximum likelihood detection}
\acro{MLE}{maximum likelihood estimator}
\acro{MMSE}{minimum mean squared error}
\acro{MRC}{maximum ratio combining}
\acro{MS}{mobile stations}
\acro{MSE}{mean squared error}
\acro{MSK}{Minimum-shift keying}
\acro{MSSS}[MSSS]	{mean-square signal separation}
\acro{MTC}{machine type communication}
\acro{MU}{multi user}
\acro{MVUE}{minimum variance unbiased estimator}
\acro{NEF}{noise enhancement factor}
\acro{NLOS}{non-line-of-sight}
\acro{NMSE}{normalized mean-squared error}
\acro{NOMA}{non-orthogonal multiple access}
\acro{NPR}{near-perfect reconstruction}
\acro{NRZ}{non-return-to-zero}
\acro{OFDM}{orthogonal frequency division multiplexing}
\acro{OFDMA}{orthogonal frequency division multiple access}
\acro{OOB}{out-of-band}
\acro{OQAM}{offset quadrature amplitude modulation}
\acro{OQPSK}{offset quadrature phase shift keying}
\acro{OTFS}{orthogonal time frequency space}
\acro{PA}{power amplifier}
\acro{PAM}{pulse amplitude modulation}
\acro{PAPR}{peak-to-average power ratio}
\acro{PC-CC}{parallel concatenated convolutional code}
\acro{PCP}{pseudo-circular pre/post-amble}
\acro{PD}{probability of detection}
\acro{pdf}{probability density function}
\acro{PDF}{probability distribution function}
\acro{PDP}{power delay profile}
\acro{PFA}{probability of false alarm}
\acro{PHY}{physical layer}
\acro{PIC}{parallel interference cancellation}
\acro{PLC}{power line communication}
\acro{PMF}{probability mass function}
\acro{PN}{pseudo noise}
\acro{ppm}{parts per million}
\acro{PRB}{physical resource block}
\acro{PRB}{physical resource block}
\acro{PSD}{power spectral density}
\acro{OAI}{OpenAirInterface}
\acro{gNB}{next generation node B}
\acro{GPIO}{general purpuse input/output}
\acro{cRIO}{CompactRIO}
\acro{SigMF}{ Signal Metadata Format}
\acro{JSON}{JavaScript object notation}
\acro{Q}{quadrature-phase}
\acro{QAM}{quadrature amplitude modulation}
\acro{QoS}{quality of service}
\acro{QPSK}{quadrature phase shift keying}
\acro{R/W}{read-or-write}
\acro{RE}{resource element}
\acro{RAM}{random-access memmory}
\acro{RAN}{radio access network}
\acro{RAT}{radio access technologies}
\acro{RC}{raised cosine}
\acro{RF}{radio frequency}
\acro{rms}{root mean square}
\acro{RRC}{radio resource control}
\acro{RW}{read-and-write}
\acro{SC}{single-carrier}
\acro{SCA}{single-carrier access}
\acro{SC-FDE}{single-carrier with frequency domain equalization}
\acro{SC-FDM}{single-carrier frequency division multiplexing}
\acro{SC-FDMA}{single-carrier frequency division multiple access}
\acro{SD}{sphere decoding}
\acro{SDD}{space-division duplexing}
\acro{SDMA}{space division multiple access}
\acro{SDR}{software-defined radio}
\acro{SDW}{software-defined waveform}
\acro{SEFDM}{spectrally efficient frequency division multiplexing}
\acro{SE-FDM}{spectrally efficient frequency division multiplexing}
\acro{SER}{symbol error rate}
\acro{SIC}{successive interference cancellation}
\acro{SINR}{signal-to-interference-plus-noise ratio}
\acro{SIR}{signal-to-interference ratio}
\acro{SISO}{single-input, single-output}
\acro{SMS}{Short Message Service}
\acro{SNR}{signal-to-noise ratio}
\acro{STC}{space-time coding}
\acro{STFT}{short-time Fourier transform}
\acro{STO}{symbol time offset}
\acro{SU}{single user}
\acro{SVD}{singular value decomposition}
\acro{TD}{time-domain}	
\acro{TDD}{time-division duplexing}
\acro{TDMA}{time-division multiple access}
\acro{TFL}{time-frequency localization}
\acro{TO}{time offset}
\acro{TS-OQAM-GFDM}{time-shifted OQAM-GFDM}
\acro{UE}{user equipment}
\acro{UFMC}{universally filtered multicarrier}
\acro{UL}{uplink}
\acro{US}{uncorrelated scattering}
\acro{USB}{universal serial bus}
\acro{UW}{unique word}
\acro{VLC}{visible light communications}
\acro{VR}{virtual reality}
\acro{WCP}{windowing and \acs{CP}}	
\acro{WHT}{Walsh-Hadamard transform}
\acro{WiMAX}{worldwide interoperability for microwave access}
\acro{WLAN}{wireless local area network}
\acro{W-OFDM}{windowed-\acs{OFDM}}	
\acro{WOLA}{windowing and overlapping}	
\acro{WSS}{wide-sense stationary}
\acro{ZCT}{Zadoff-Chu transform}
\acro{ZF}{zero-forcing}
\acro{ZMCSCG}{zero-mean circularly-symmetric complex Gaussian}
\acro{ZP}{zero-padding}
\acro{ZT}{zero-tail}
\acro{URLLC}{ultra-reliable low-latency communications}

\acro{HSI}{human system interface}
\acro{HMI}{human machine interface}
\acro{VR} {visual reality} 
\acro{AGV}{automated guided vehicles}
\acro{MEC}{multiaccess edge cloud}
\acro{TI} {tactile Internet}
\acro{IMT}{ international mobile telecommunications}
\acro{GN}{gateway node}
\acro{CN}{control node}
\acro{NC}{network controller}
\acro{SN}{sensor node}
\acro{AN}{actuator node}
\acro{HN}{haptic node}
\acro{TD}{tactile devices}
\acro{SE}{spectral efficiency}
\acro{AI}{artificial intelligence}
\acro{TSM}{tactile service manager}
\acro{TTI}{transmission time interval}
\acro{NR}{new radio}
\acro{SDN}{software defined networking}
\acro{NFV}{ network function virtualization}
\acro{CPS}{cyber-physical system}
\acro{TSN}{Time-Sensitive Networking}
\acro{FEC}{forward error correction}
\acro{STC}{space-time  coding}
\acro{HARQ}{hybrid automatic repeat request}
\acro{CoMP} {Coordinated multipoint}
\acro{HIS}{human system interface }
\acro{RU}{radio unit}
\acro{CU}{central unit}
\acro{VoIP}{voice over IP}
\acro{B5G}{byeond 5G}
\acro{KPI}{key performance indicator}
\acro{ERP}{effective radiated power}
\acro{ZXM}{zero crossing modulation}
\acro{PAE}{power added efficiency}
\acro{AGC}{automatic gain control}
\acro{NTN}{non-terrestrial networks}
\acro{UPA}{uniform planner array}
\acro{MRT}{maximum ratio transmission}
\end{acronym}
	\title{Impact of Phase Errors on \\Distributed NTN Beam Focusing}
    \author{
      \IEEEauthorblockN{
        Ahmad Nimr, Mohammad Parvini, Bitan Banerjee, 
        Gerhard Fettweis
      }
      \IEEEauthorblockA{
        Vodafone Chair Mobile Communication Systems, Technische Universit\"{a}t Dresden, Dresden, Germany\\
        Email: \{ahmad.nimr, mohammad.parvini, bitan.banerjee, gerhard.fettweis\}@tu-dresden.de
      }
    }
	
	\maketitle
	\IEEEpeerreviewmaketitle
	\begin{abstract}
	This paper investigates distributed beam focusing for coordinated satellite constellations with phased arrays, motivated by future non-terrestrial network (NTN) systems. A geometric and channel model is developed by incorporating satellite positions, array orientations, antenna directivity, and polarization effects. Under ideal synchronization, the achievable coherent combining gain is analyzed for different constellation geometries, showing that maximum ratio transmission (MRT) enables quadratic scaling of the received power with the number of satellites.
	The impact of phase errors caused by residual synchronization, timing, mobility, and localization mismatches is then investigated. Closed-form expressions for the average coherent gain are derived for uniformly distributed timing offsets, demonstrating the transition from coherent to non-coherent combining. The results show that synchronization and timing mismatches reduce the coherent combining gain, while geometry-dependent effects govern the resulting spatial focusing behavior.
	Numerical results further show that linear and circular constellations provide different focusing characteristics and spatial separation capabilities. However, MRT-based focusing results in strong sidelobes and limited spatial division capability, motivating the need for joint analog beamforming and digital precoding optimization to improve spatial selectivity and robustness against mobility and localization errors.
\end{abstract}
\begin{IEEEkeywords}
	Distributed NTN, distributed beam focusing, coherent transmission, phase errors, synchronization, timing offsets, antenna arrays.
\end{IEEEkeywords}

	\acresetall
\section{Introduction}\label{sec:introduction}

Distributed beamforming enables multiple spatially separated transmitters to coherently focus electromagnetic energy toward a target user through coordinated phase and amplitude control \cite{Mudumbai2009DistributedBeamforming, DistributedPhasedArrays2021}. Compared with conventional co-located massive \ac{MIMO} systems, distributed architectures provide significantly larger effective aperture, improved spatial diversity, and more flexible coverage. Such concepts are particularly relevant for future \acp{NTN} \cite{guidotti_etal_19, Kodheli2021SatelliteSurvey, Azari2022NTNSurvey}, where multiple satellites equipped with phased arrays may jointly serve users on the ground through coordinated transmission. Similar distributed coherent transmission concepts also appear in coordinated terrestrial access points, distributed radar, integrated sensing and communications, and cell-free massive \ac{MIMO} systems \cite{Ngo2017CellFree, Interdonato2019CFMassiveMIMO, Meng2025NearFieldISAC}.

The fundamental benefit of distributed coherent transmission is the ability to coherently combine signals from multiple transmitters at the intended receiver. As shown later in this paper, under ideal synchronization and perfect geometry knowledge, the received signal power scales quadratically with the number of coherent transmitters. However, achieving such gains requires accurate phase alignment across distributed transmitters, which depends on synchronization accuracy, propagation delay compensation, geometry knowledge, and array calibration \cite{DistributedPhasedArrays2021,Brown2008DistributedCarrierSync}. Consequently, distributed beamforming is often considered highly sensitive to residual timing and carrier phase mismatches \cite{banerjeeNTN2026Conf}.

Practical systems can typically estimate and compensate synchronization offsets with high accuracy using pilot-based synchronization and tracking techniques \cite{Tutzi24Dist}. In contrast, geometry-dependent effects remain more difficult to mitigate. Residual localization errors, mobility, and propagation delay variations directly translate into phase mismatches that degrade coherent combining. Furthermore, unlike co-located arrays, distributed constellations experience different propagation directions and array orientations across transmitters, making the achievable focusing behavior strongly dependent on geometry. In multi-user scenarios, these effects become even more critical since conventional \ac{MRT}-based focusing provides limited spatial selectivity within the covered area \cite{MicroSynchDistMIMO,Lu2024NearFieldTutorial,An2024NearFieldCommunications,Banerjee26}.

Motivated by these observations, this paper investigates distributed beam focusing by developing a geometric and channel model that explicitly incorporates satellite positions, array orientations, antenna directivity, and polarization effects. A statistical framework is then developed to analyze the impact of residual phase errors caused by synchronization mismatch, timing offsets, mobility, and localization uncertainty. Closed-form expressions are derived for the average coherent gain under uniformly distributed timing offsets, demonstrating the transition from coherent to non-coherent combining. Numerical results further illustrate how different constellation geometries provide different focusing characteristics and spatial separation capabilities. The analysis also highlights the limitations of \ac{MRT}-based focusing in mobility and multi-user scenarios, motivating the need for joint analog beamforming and digital precoding optimization. The main contributions of this paper are summarized as follows:
\begin{itemize}
    \item First, a geometry-aware distributed NTN channel model is developed that explicitly incorporates satellite positions, array orientations, antenna directivity, and polarization effects. 
    \item Second, the achievable coherent focusing gain under ideal synchronization is analyzed, demonstrating the quadratic scaling behavior of distributed coherent transmission. 
    \item Third, a statistical framework is developed to characterize synchronization-, timing-, and localization-induced phase errors, and closed-form expressions are derived for the average coherent combining gain. 
    \item And finally, numerical results are presented to evaluate different satellite constellation geometries (linear and circular \cite{parvinilinearcircular}) and to investigate their corresponding spatial focusing and user-separation characteristics.
\end{itemize}

The remainder of this paper is organized as follows. \secref{sec:section2} presents the system and channel model. \secref{sec:section3} analyzes ideal distributed beam focusing and multi-user operation. \secref{sec:section4} presents the numerical results and investigates the impact of synchronization and timing errors. Finally, concluding remarks are provided in \secref{sec:conclusions}.
	\section{System model}\label{sec:section2}
\begin{figure}[t]
	\vspace{0.05in}
	\centering	\includegraphics[width=.46\textwidth]{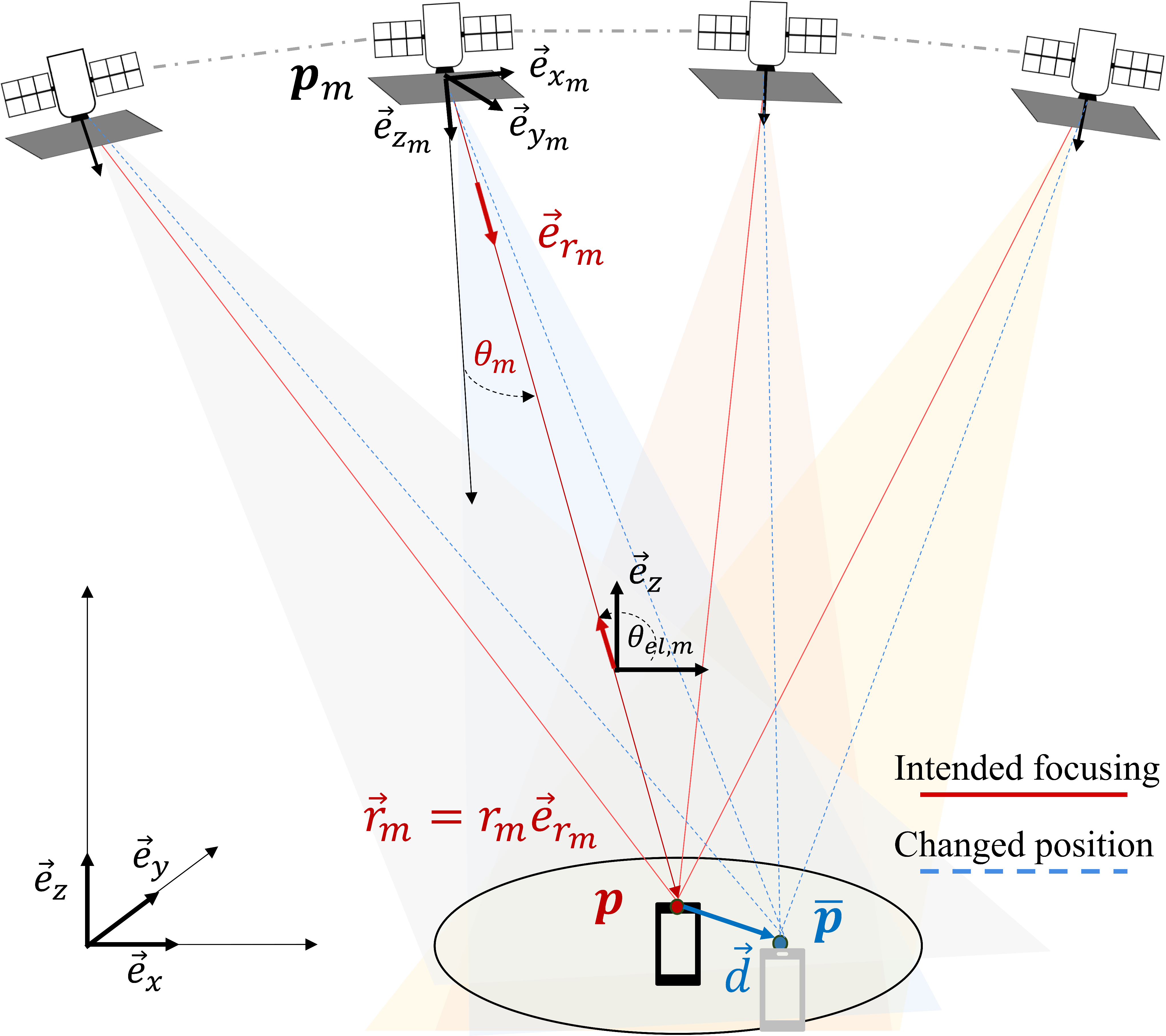}
	\caption{Distributed \ac{NTN} beam focusing geometry and system model }\label{fig:system_model}
	\vspace{-10pt}
\end{figure}

Consider a distributed \ac{NTN} consisting of $M$ satellites. Each satellite is equipped with a planar phased array centered at position $\ma{p}_m$ in a global coordinate system, as illustrated in \figref{fig:system_model}.
 The antenna elements are assumed to have identical radiation patterns that are approximately constant over the region of interest. The satellites are interconnected and can be centrally coordinated by a dedicated processing unit, which may be co-located with one of the satellites.  The objective is to jointly serve ground users by configuring both analogue (per-element) and digital (per-array) phase and gain.
Assuming that each satellite array is driven by a single \ac{RF} chain, the distributed beamforming reduces to phase and gain control at the array level combined with digital precoding across satellites. This architecture limits the spatial degrees of freedom in multi-user scenarios. 
Users are equipped with a single antenna with approximately isotropic response toward the satellite constellation. The received signal is given by the superposition of the signals transmitted from all satellites.
\subsection{Geometry and coordinate system}
A global coordinate system is defined by the orthonormal basis $\{\vec{e}_x, \vec{e}_y, \vec{e}_z\}$ with origin at $[0,0,0]^{\text{T}}$. All positions are expressed in this coordinate system. The local coordinate system of the $m$-th array is defined as
\begin{equation}
	\{\vec{e}_{x_m}, \vec{e}_{y_m}, \vec{e}_{z_m}\}
	=
	\{\vec{e}_x, \vec{e}_y, \vec{e}_z\}\ma{\Omega}_m,
\end{equation}
where $\ma{\Omega}_m$ is a unitary orientation matrix with origin at $\ma{p}_m$. The orientation $\ma{\Omega}_m$ captures the rotation of the array with respect to the global coordinate system and is assumed to be known.
The vector from the $m$-th array to a point $\ma{p}$ is
\begin{equation}
	\vec{r}_m = \ma{p} - \ma{p}_m = r_m \vec{e}_{r_m}.
\end{equation}
The corresponding angular representation is given by
\begin{equation}
	\vec{e}_{r_m}
	=
	\sin \theta_m
	\left(
	\cos \phi_m \vec{e}_{x_m}
	+
	\sin \phi_m \vec{e}_{y_m}
	\right)
	+
	\cos \theta_m \vec{e}_{z_m}.
\end{equation}
The angle $\theta_m$ is defined with respect to the local array normal $\vec e_{z_m}$, following the conventional array-processing notation.
The satellite elevation angel $\theta_{\mathrm{el},m}$ is defined with respect to the global horizontal plane. Assuming that $\vec e_z$ denotes the upward direction, the elevation angle is given by
\begin{equation}
	\sin \theta_{\mathrm{el},m}
	=
	-\vec{e}_{r_m}\cdot\vec e_z.
\end{equation}
\subsection{Array response and channel model}
Consider a \ac{UPA} of size $N_x \times N_y$, with element spacing $\Delta = \frac{\lambda_c}{2}$, where $\lambda_c = \frac{c}{f_c}$ is the wavelength at carrier frequency $f_c$, and $c$ is the speed of light. The element positions are given by
\begin{equation*}
	\ma{p}_{m,n_x,n_y} = \ma{p}_m + \vec{\Delta}_{m,n_x,n_y},
\end{equation*}
where
\begin{equation*}
	\vec{\Delta}_{m,n_x,n_y} = \left(n_x - \tfrac{N_x-1}{2}\right)\Delta \vec{e}_{x_m} + \left(n_y - \tfrac{N_y-1}{2}\right)\Delta \vec{e}_{y_m},
\end{equation*}
for $n_x = 0,\dots,N_x-1$ and $n_y = 0,\dots,N_y-1$.
The array aperture is $D = \Delta \sqrt{N_x^2 + N_y^2}$. For $r_m \gg \frac{2D^2}{\lambda_c}$, the point $\ma{p}$ lies in the far-field of the array, and
\begin{equation}
	\Norm{\ma{p} - \ma{p}_{m,n_x,n_y}} \approx r_m - \vec{e}_{r_m} \cdot \vec{\Delta}_{m,n_x,n_y}.
\end{equation}
The corresponding array response for $f\in [-\frac{B}{2}, \frac{B}{2}]$, where $B$ is the bandwidth, is given by
\begin{equation}
	A_m(f+f_c) = \sum_{n_x,n_y} w_{m,n_x,n_y} 
	e^{j2\pi \left(1+\frac{f}{f_c}\right)\varphi_{m,n_x,n_y}},
\end{equation}
where  $w_{m, n_x, n_y}$ is the complex phase and gain control, and 
\begin{equation}
	\begin{aligned}
		\varphi_{m,n_x,n_y}
		&= \frac{f_c}{c} \vec{e}_{r_m} \cdot \vec{\Delta}_{m,n_x,n_y} \\
		&= \frac{1}{2}\left(n_x - \tfrac{N_x-1}{2}\right)\sin\theta_m \cos\phi_m \\
		&\quad + \frac{1}{2}\left(n_y - \tfrac{N_y-1}{2}\right)\sin\theta_m \sin\phi_m.
	\end{aligned}
\end{equation}
The frequency-domain \ac{LoS} channel between the user and the $m$-th array is
\begin{equation}
	\tilde{h}_m(f) = \alpha_m(f) A_m(f+f_c) \frac{e^{-j2\pi (f+f_c)\frac{r_m}{c}}}{r_m},
\end{equation}
where $\alpha_m(f)$ captures antenna directivity and polarization effects.
Under the narrowband condition \cite{parviniWxlmimo}
\begin{equation}
	|B| \ll \frac{f_c}{\sqrt{N_x^2 + N_y^2}},
\end{equation}
the beam squint is negligible, and
\begin{equation}
	A_m(f+f_c) \approx A_m(f_c) = \sum_{n_x,n_y} w_{m,n_x,n_y} e^{j2\pi \varphi_{m,n_x,n_y}}.
\end{equation}
Moreover, $\alpha_m(f) \approx \alpha_m$ and its magnitude can be related to the transmit and receive antenna gains as
\begin{equation}
	|\alpha_m|^2 = \frac{\lambda_c^2}{(4\pi)^2} D_{\text{tx}}(\theta_m,\phi_m) D_{\text{rx}}(\theta_m,\phi_m),
\end{equation}
where $D_{\text{tx}}(\cdot)$ and $D_{\text{rx}}(\cdot)$ are the antenna  directivities. The resulting narrowband channel is
\begin{equation}
	\tilde{h}_m(f) = \frac{\alpha_m A_m}{r_m} e^{-j2\pi (f+f_c)\tau_m}, \label{eq:channel}
\end{equation}
where $\tau_m = \frac{r_m}{c}$. The channel can be controlled via the array response $A_m$ by appropriately selecting the weights $w_{m,n_x,n_y}$ in relation to $\varphi_{m,n_x,n_y}$, which depends on the user location as well as the array position and orientation. Moreover, $\alpha_m$ is an antenna-related parameter that captures the effective response of the transmit antenna elements and the receive antenna, including directivity and polarization. Furthermore, $r_m$ and $\tau_m$ are purely geometrical parameters determined by the satellite and user locations.

	\section{Ideal distributed beam focusing}\label{sec:section3}
In this analysis, it is assumed that all satellites share a common time and frequency reference, and that all information related to user location, array positions, and orientations is perfectly known.
\subsection{Single user}
Consider a user located at position $\ma{p}$. All satellites transmit a common signal $x(t)$ and jointly select their array beamforming weights and digital precoding coefficients $\gamma_m$ to maximize the received energy at the user of interest.The received signal in the frequency domain is
\begin{equation}
	\tilde{y}(f) = \sum_{m=1}^{M} \gamma_m \tilde{h}_m(f) \tilde{x}(f) + \tilde{v}(f),
\end{equation}
where $\tilde{v}(f)$ is additive noise. A per-satellite power constraint implies $|\gamma_m| \leq 1$. The coefficients are selected to maximize the received power, leading to
\begin{align}
	\max_{\{\gamma_m\}} G = \left| \sum_{m=1}^{M} \gamma_m \frac{\alpha_m A_m}{r_m} e^{-j2\pi (f+f_c)\tau_m} \right|^2.
\end{align}

The optimal solution follows \ac{MRT}, given by
\begin{equation}
	\gamma_m = \frac{(\alpha_m A_m)^*}{|\alpha_m A_m|} e^{j2\pi (f+f_c)\tau_m}.
\end{equation}
The resulting coherent gain is
\begin{equation}
	G_{\text{max}} = \left| \sum_{m=1}^{M} \frac{|\alpha_m| |A_m|}{r_m} \right|^2. \label{eq:max_gain}
\end{equation}
Here, $|A_m|$ is maximized by choosing
\begin{equation}
	w_{m,n_x,n_y} = e^{-j2\pi \varphi_{m,n_x,n_y}},
\end{equation} 
leading to $A_m = A = N_xN_y$, while  $\alpha_m$ depends on the relative orientation and polarization alignment between the transmit arrays and the user antenna. While the transmit-side characteristics are determined by the satellite design, the effective response is also influenced by the user device orientation, which is generally not controlled by the network.
The combined weights $\gamma_m A_m$ correspond to beam focusing, where $A_m$ depends on the direction, while $\gamma_m$ compensates for propagation delays through $\tau_m$.
Assuming $r_m \approx r$ and $\alpha_m \approx \alpha$, the maximum gain becomes
\begin{equation}
	G_{\text{max}} = M^2 \frac{|\alpha|^2}{r^2}A^2.
\end{equation}
This result shows a quadratic scaling  of the link budget of individual arrays $\frac{|\alpha|^2}{r^2}A^2$,  highlighting the potential of distributed coherent transmission. Achieving this gain requires accurate location information, precise array calibration, and ideal synchronization across satellites.
\subsection{Focus sensitivity with mobility}
If the user moves within the coverage area to a new position $\bar{\ma{p}} = \ma{p} + \vec{d}$, where $\vec{d}$ is a displacement vector, the variation in direction with respect to the arrays is negligible, while the delay variation becomes
\begin{equation}
	\Delta{\tau_m} = \frac{1}{c} \left(\Norm{\ma{p} - \ma{p}_m} - \Norm{\bar{\ma{p}} - \ma{p}_m} \right)\approx -\frac{1}{c}\vec{e}_{r_m} \cdot \vec{d}.
\end{equation}
Without updating the digital precoding to the new position, the resulting gain is
\begin{equation}
	G(\vec{d}) = 
	\frac{|A \alpha|^2}{r^2}  \left| \sum_{m=1}^{M}e^{j2\pi (f+f_c) \Delta{\tau_m}} \right|^2. \label{eq:gain_with_mobility}
\end{equation}
This indicates that user mobility introduces phase misalignment across satellites.  Consequently, the analog beamforming design needs to account for mobility, which reduces the achievable peak gain at the focal point.
This is similar to non-coherent combining gain ($\alpha_m = 1$), where 
\begin{equation}
		G_{\text{NC}} = 
		\frac{|A \alpha|^2}{r^2}  \left| \sum_{m=1}^{M}e^{-j2\pi (f+f_c) {\tau_m}} \right|^2. \label{eq:non-coherent}
\end{equation}
Similar effect arises in the case of user localization error, which leads to phase error, as discussed in \secref{sec:phase erorr}.
\subsection{Multiple users}
Consider $K < M$ users. The channel between the $k$-th user and the $m$-th array is given by
\begin{equation}
	\tilde{h}_{k,m}(f) = \frac{\alpha_{k,m} A_{k,m}}{r_{k,m}} e^{-j2\pi (f+f_c)\tau_{k,m}},
\end{equation}
where $r_{k,m} = \Norm{\ma{p}_k - \ma{p}_m}$, $A_{k,m}$ is the array gain in the direction of the $k$-th user, and $\alpha_{k,m}$ captures the antenna response.
Serving multiple users requires joint analog beamforming and digital precoding. The analog coefficients $\{w_{m,n_x,n_y}\}$ determine the array response $A_{k,m}$ toward each user, while the digital precoding matrix $\ma{\Gamma} \in \compl^{M \times K}$ manages inter-user interference. The received signal at the $k$-th user is
\begin{equation}
	\tilde{y}_k(f) = \tilde{\ma{h}}_{k}^T(f)\ma{\Gamma}\tilde{\ma{x}}(f) + \tilde{v}_k(f),
\end{equation}
where  $\tilde{\ma{h}}_{k}^T(f) = [\tilde{h}_{k,1}(f), \cdots, \tilde{h}_{k,M}(f)]$, and $\tilde{\ma{x}}(f) = [\tilde{x}_1(f),\cdots, \tilde{x}_K(f)]^{\text{T}}$.
The performance depends on the ability of the analog beamforming to create sufficiently distinct channels across users, which is limited by the array geometry and the antenna response.

In a special case, the arrays are steered toward a reference user ($k=1$) to achieve the maximum coherent gain as in \eqref{eq:max_gain}. As a result, the array response is maximized in this direction, i.e., $A_m \approx A$. For users located within the main lobe of the arrays, the variation in the array response is small, and thus $A_{k,m} \approx A$, $\alpha_{k,m} \approx \alpha$, and $r_{k,m} \approx r$. The channel can then be approximated as
\begin{equation}
	\tilde{h}_{k,m}(f) \approx \frac{\alpha A}{r} e^{-j2\pi (f+f_c)\tau_{k,m}}.
\end{equation}
For a displacement $\vec{d}_k$ from the reference user,
\begin{equation}
	\Delta{\tau_{m,k}} \approx -\frac{1}{c}\vec{e}_{r_m} \cdot \vec{d}_k, \quad k>1,
\end{equation}
where $\tau_m$ corresponds to the delay of the reference user.
Using matched filtering, the interference from the reference user to the $k$-th user is characterized by
\begin{equation}
	G(\vec{d}_k) = 
	\frac{|A \alpha|^2}{r^2}
	\left| \sum_{m=1}^{M} e^{j2\pi (f+f_c) \Delta{\tau_{m,k}}} \right|^2.
\end{equation}
This shows that users within the same beam experience similar array gain, while the residual delay differences limit interference suppression. As a result, \ac{MRT}-based focusing alone limits spatial multiplexing capability in distributed beamforming systems.

	\section{Impact of synchronization errors}\label{sec:section5}
Consider residual time offset $\Delta t_m$, frequency offset $\Delta f_m$, and phase offset $\Delta \varphi_m$ of the $m$-th array relative to the reference array $m=1$.
The transmitted baseband signal at the $m$-th array is \cite{mengalisynch}
\begin{equation*}
	x_m(t) = x(t-\Delta t_m) e^{j2\pi \Delta f_m (t-\Delta t_m)} e^{-j2\pi f_c\Delta t_m} e^{j\Delta \varphi_m}.
\end{equation*}
In the frequency domain,
\begin{equation*}
	\tilde{x}_m(f) = e^{j\Delta \varphi_m} \tilde{x}(f-\Delta f_m) e^{-j2\pi (f+f_c)\Delta t_m}.
\end{equation*}
For small frequency offsets, $|\Delta f_m t| \ll 1$ during one symbol duration, using the approximation $e^{j2\pi \Delta f_m t} \approx 1+ j2\pi  \Delta f_m t $ results in  
\begin{equation}
	\tilde{x}_m(f) \approx \tilde{x}(f) e^{j\Delta \varphi_m} e^{-j2\pi (f+f_c)\Delta t_m} + \tilde{z}_m(f),
\end{equation}
where $\tilde{z}_m(f)$ represents a self-interference term due to frequency offset given by 
\begin{equation*}
	\tilde{z}_m(f) \approx j2\pi \Delta f_m e^{j\Delta \varphi_m} e^{-j2\pi (f+f_c)\Delta t_m} \int t\, x(t) e^{-j2\pi f t} dt.
\end{equation*}
Assuming that beamforming and digital precoding are designed for ideal synchronization, the received signal becomes
\begin{equation}
	\begin{split}
		\tilde{y}(f) &= \frac{|\alpha A|}{r} \tilde{x}(f) \sum_{m=1}^{M} e^{j\Delta \varphi_m}  e^{-j2\pi (f+f_c)\Delta t_m}\\
		& + \frac{|\alpha A|}{r}\sum_{m=1}^{M} \tilde{z}_m(f) + \tilde{v}(f),
	\end{split}
\end{equation} 
The resulting \ac{SNR} is
\begin{equation}
	\rho = 
	\frac{\mathbb{E}\!\left[\left|\sum_{m=1}^{M} e^{j\Delta \varphi_m} e^{-j2\pi (f+f_c)\Delta t_m}\right|^2\right]}
	{\mathbb{E}\!\left[\left|\sum_{m=1}^{M} \tilde{z}_m(f)\right|^2\right] + \frac{r^2}{|\alpha A|^2} \sigma^2}. \label{eq:SNR}
\end{equation}
The residual time and phase offsets reduce the coherent combining gain, similarly to the effect of mobility under ideal synchronization and non-coherent combining \eqref{eq:non-coherent}. In contrast, the residual frequency offset introduces an additional interference term, which is typically less dominant for small offsets.
\subsection{Properties of phase errors}\label{sec:phase erorr}
The coherent gain depends on the phase error terms  through
\begin{equation}
	G_d = \left| \sum_{m=1}^{M} e^{j\Phi_m} \right|^2,
\end{equation}
where 	$\Phi_m = -2\pi (f+f_c)\Delta t_m + \Delta \varphi_m$ is a random variable that captures the residual phase mismatch due to synchronization and timing errors \eqref{eq:SNR}, mobility-induced phase shifts or localization error \eqref{eq:gain_with_mobility}, or other non-coherent effects. By expressing $e^{j\Phi_m}$ as a vector,
\begin{equation}
	\ma{s}_m = \left[\cos \Phi_m, \sin \Phi_m \right]^T,
\end{equation}
with $\Norm{\ma{s}_m}^2 = 1$. Assuming identical and independent phase errors across the satellites, the mean vector and covariance matrix are
\begin{equation}
	\begin{split}
		\ma{\mu} &= \Ex{\ma{s}_m} = \left[\mu_1, \mu_2 \right]^T,\\
		\ma{C} &= \Ex{(\ma{s}_m-\ma{\mu})(\ma{s}_m-\ma{\mu})^T}.
	\end{split} \label{eq:mean_var}
\end{equation}

The average coherent gain is
\begin{equation}
	\Ex{G_d} 
	= \Ex{\Norm{\sum_{m=1}^{M} \ma{s}_m}^2}
	= M + M(M-1)\Norm{\ma{\mu}}^2.
\end{equation}
This shows that the coherent gain depends on $\Norm{\ma{\mu}}^2$, where $\Norm{\ma{\mu}}^2 \approx 1$ corresponds to near-coherent combining (${\Ex{G_d}\approx M^2}$), while $\Norm{\ma{\mu}}^2 \approx 0$ corresponds to non-coherent combining ($\Ex{G_d}\approx M$).

For sufficiently large $M$, the sum $\ma{s} = \sum_{m=1}^{M} \ma{s}_m$
can be approximated by a Gaussian random vector as
\begin{equation}
	\ma{s} \overset{\textrm{approx.}}{\sim} \mathcal{N}\left(M\ma{\mu}, M\ma{C}\right).
\end{equation}
Since $G_d = \Norm{\ma{s}}^2$, the distribution of $G_d$ follows a generalized non-central chi-square distribution, 

\subsection{Impact of timing offsets}
The phase error $\Phi_m$ is highly influenced by the residual timing error $\Delta t_m$, whereas the constant phase offset $\Delta \varphi_m$ can be calibrated and compensated. In practical systems, timing offsets can be estimated and corrected with sub-sample accuracy, such that the residual timing error satisfies $|\Delta t_m| \ll 1/B$. As a result, the dominant contribution becomes $\Phi_m \approx -2\pi f_c \Delta t_m$. Therefore, the impact of timing errors depends on the fractional part of $f_c \Delta t_m$.

\begin{figure}[t]
	\vspace{0.05in}
	\centering
	
	\subfloat[Mean gain]{
			\includegraphics[width=.23\textwidth]{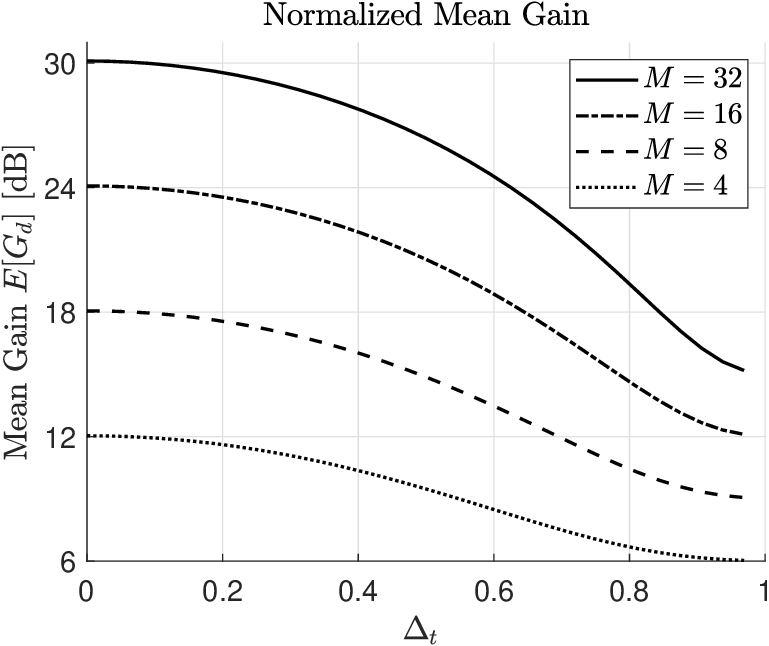}
		\label{fig:mean_gain}
	}
	\subfloat[\ac{CCDF} of the gain]{
		\includegraphics[width=.23\textwidth]{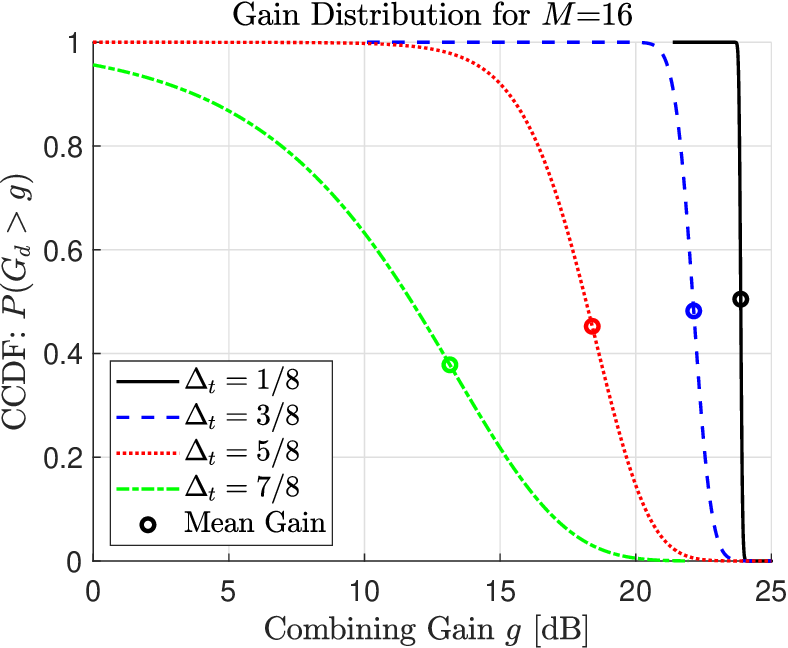}
		\label{fig:gain_distribution_M_16}
	}
	\caption{Statistical coherent combining gain under phase error for different values of $M$ and $\Delta_t$}
	\label{fig:gain}
	\vspace{-10pt}
\end{figure}

Assume that the fractional part of $f_c \Delta t_m$ is uniformly distributed in $[-\frac{\Delta_t}{2}, \frac{\Delta_t}{2}]$, where $0 \leq \Delta_t \leq 1$.  Then, the phase error $\Phi_m$ is uniformly distributed with
\begin{equation}
	f_{\Phi_m} (\Phi) = \frac{1}{2\pi \Delta_t}, 
	\quad 
	\Phi \in [-\pi \Delta_t, \pi \Delta_t].
\end{equation}
Accordingly, the mean vector and covariance matrix in \eqref{eq:mean_var} are given by
\begin{equation}
	\ma{\mu}
	=
	\begin{bmatrix}
		\frac{\sin(\pi\Delta_t)}{\pi\Delta_t}\\
		0
	\end{bmatrix},
	\qquad
	\ma{C}
	=
	\begin{bmatrix}
		\sigma_1^2 & 0\\
		0 & \sigma_2^2
	\end{bmatrix},
\end{equation}
where
\begin{equation}
	\begin{split}
		\sigma_1^2 &= \frac{1}{2} + \frac{\sin(2\pi\Delta_t)}{4\pi\Delta_t} - \left(
		\frac{\sin(\pi\Delta_t)}{\pi\Delta_t} \right)^2,\\
		\sigma_2^2	&= \frac{1}{2} - \frac{\sin(2\pi\Delta_t)}{4\pi\Delta_t}.
	\end{split}
\end{equation}
Therefore,
\begin{equation}
	\small
	\Ex{G_d} = M^2 \left( \frac{\sin(\pi\Delta_t)}{\pi\Delta_t} \right)^2 + M \left[ 1- \left( \frac{\sin(\pi\Delta_t)}{\pi\Delta_t} \right)^2 \right].
\end{equation}
This result shows that timing offsets mainly introduce a gradual reduction in coherent gain, as illustrated in \figref{fig:mean_gain}. For small $\Delta_t$, the gain remains close to the coherent scaling $M^2$, while large timing uncertainty leads to non-coherent combining with gain scaling approximately proportional to $M$. The \ac{CCDF} of the combining gain is illustrated in \figref{fig:gain_distribution_M_16} for $M=16$. It can be observed that a phase error smaller than approximately $\frac{\pi}{8}$ ($\Delta_t=\frac{1}{8}$) is required to maintain the achieved gain close to the maximum coherent gain $M^2$.

Note that localization errors under ideal synchronization, as in \eqref{eq:gain_with_mobility}, can be analyzed similarly by considering the distribution of the fractional part of $f_c\Delta\tau_m$. Accordingly, the localization error should remain smaller than approximately $\frac{\lambda_c}{8}$ in order to maintain near-coherent focusing gain. Therefore, the overall phase error resulting from both synchronization and localization mismatches should remain below approximately $\frac{\pi}{8}$ under \ac{MRT}-based focusing.

%

	\section{Numerical results}\label{sec:section4}

This section presents numerical results to illustrate the behavior of distributed beam focusing under practical \ac{LEO} geometries. A quasi-static scenario is considered, where transmission occurs over a sufficiently short interval such that Doppler effects can be neglected. The positions of the user and satellites are assumed to be perfectly known in order to isolate the impact of geometry and phase synchronization.
\subsection{Constellation scenarios}
A \ac{LEO} constellation with altitude $a=600~\mathrm{km}$, carrier frequency $f_c=2~\mathrm{GHz}$, bandwidth $B=20~\mathrm{MHz}$, and $M=16$ satellites is considered. Each satellite is equipped with a $32\times32$ \ac{UPA} with element spacing $\Delta=\lambda_c/2$.

\begin{figure}[t]
	\vspace{0.05in}
	\centering
	\includegraphics[width=.25\textwidth]{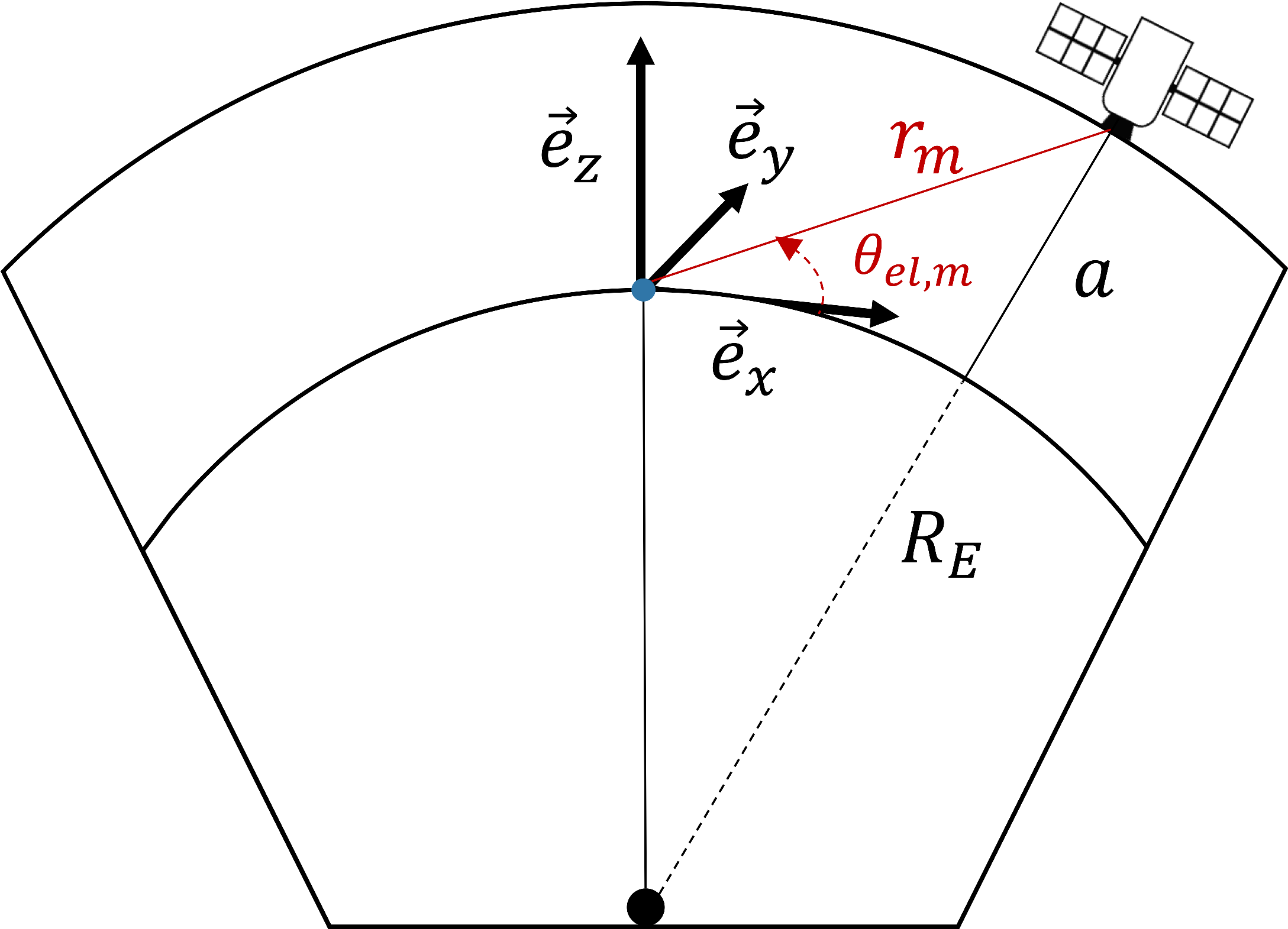}
	\caption{Geometry of the satellite elevation angle $\theta_{\mathrm{el},m}$ and the corresponding slant range $r_m$ in the considered NTN scenario}
	\label{fig:sat_position}
	\vspace{-10pt}
\end{figure}

The satellite position is determined from the slant range $r_m$, which depends on the satellite altitude $a$, Earth radius $R_E = 6371~\mathrm{km}$, and elevation angle $\theta_{\mathrm{el},m}$ as illustrated in \figref{fig:sat_position},
\begin{equation}
	r_m =
	\sqrt{
		R_E^2\sin^2\theta_{\mathrm{el},m}
		+
		a^2
		+
		2aR_E
	}
	-
	R_E\sin\theta_{\mathrm{el},m}.
\end{equation}

Together with the azimuth angle $\phi_{\mathrm{el},m}$, the satellite position relative to the user position $\ma{p}$ is given by
\begin{equation}
	\ma{p}_m
	=
	\ma{p}
	+
	r_m
	\begin{bmatrix}
		\cos\theta_{\mathrm{el},m}\cos\phi_{\mathrm{el},m}\\
		\cos\theta_{\mathrm{el},m}\sin\phi_{\mathrm{el},m}\\
		\sin\theta_{\mathrm{el},m}
	\end{bmatrix}.
\end{equation}

\begin{figure}[t]
	\vspace{0.05in}
	\centering
	\includegraphics[width=.47\textwidth]{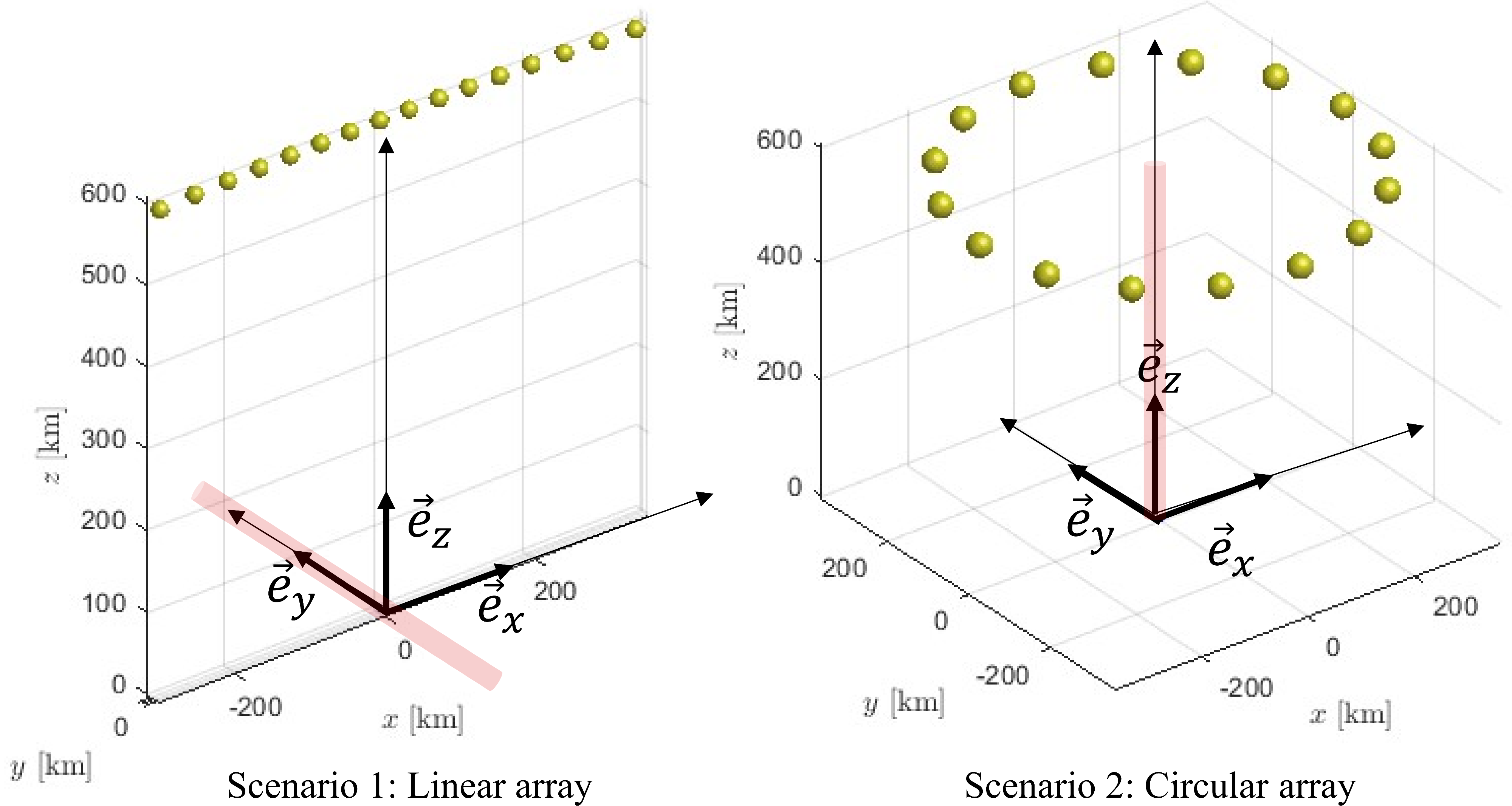}
	\caption{Considered distributed satellite geometries and the corresponding focusing characteristics}
	\label{fig:Scenarios}
	\vspace{-10pt}
\end{figure}

Two constellation geometries are considered and illustrated in \figref{fig:Scenarios}.
In the first scenario, the satellites are distributed over two opposite angular sectors along the $x$-axis. Half of the satellites are located with azimuth angle $\phi_{\mathrm{el},m}=0$, while the remaining satellites are located at $\phi_{\mathrm{el},m}=\pi$. The elevation angles are uniformly distributed between $60^\circ$ and $90^\circ$. 
%
This configuration forms a symmetric two-dimensional distributed aperture surrounding the user, enabling improved spatial localization in the $XY$ plane while extending the focal region along the $z$-axis.
\subsection{Focus evaluation}
\begin{figure}[t]
	\vspace{0.05in}
	\centering
	
	\subfloat[Gain in the $XY$ plane.]{
		\includegraphics[width=0.23\textwidth]{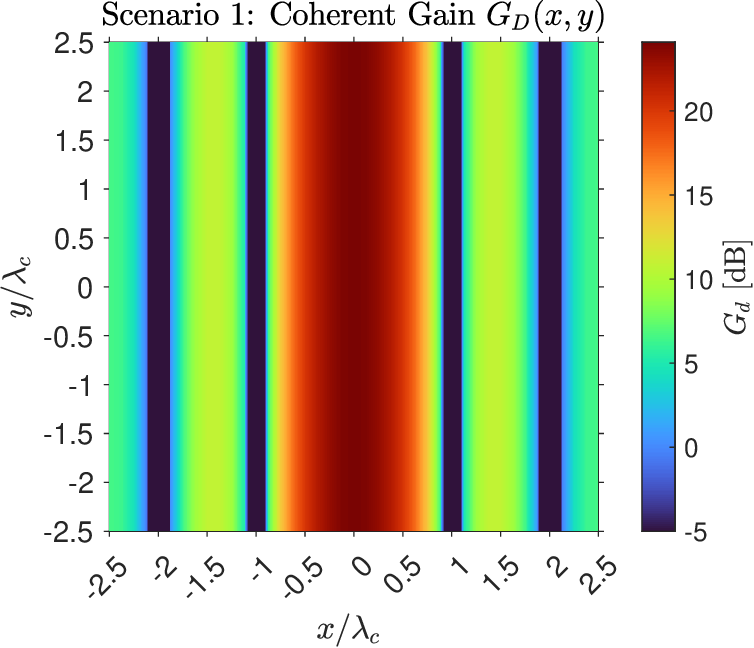}
		\label{fig:scen_1_C_1_XY_1}
	}
	\subfloat[Gain in the $YZ$ plane.]{
		\includegraphics[width=0.23\textwidth]{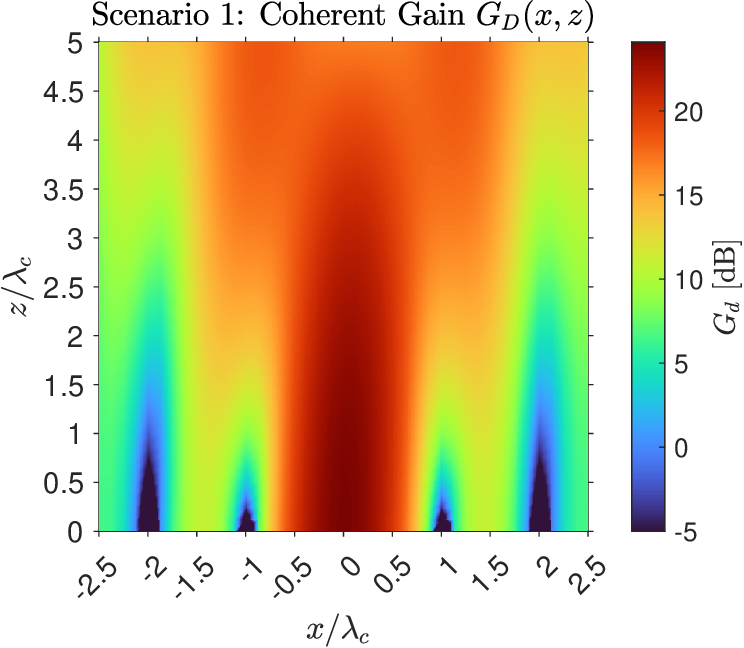}
		\label{fig:scen_1_C_1_XY_0}
	}
	\caption{Coherent combining gain of linear constellation at $\ma{p}=0$ and the resulting gain distribution in different planes.}
	\label{fig:scen_1_c}
	\vspace{-10pt}
\end{figure}

\begin{figure}[t]
	\vspace{0.05in}
	\centering
	
	\subfloat[Gain in the $XY$ plane]{
		\includegraphics[width=0.23\textwidth]{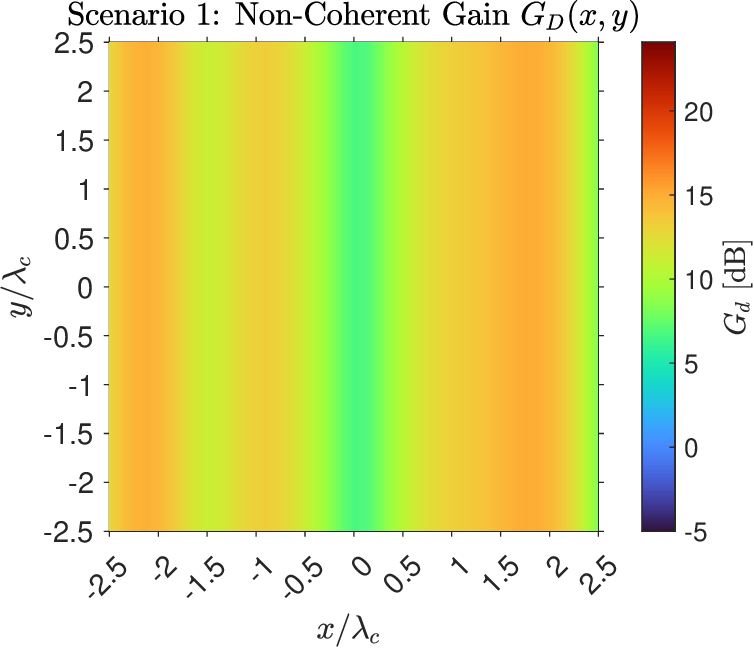}
		\label{fig:scen_1_C_0_XY_1}
	}
	\subfloat[Gain in the $YZ$ plane]{
		\includegraphics[width=0.23\textwidth]{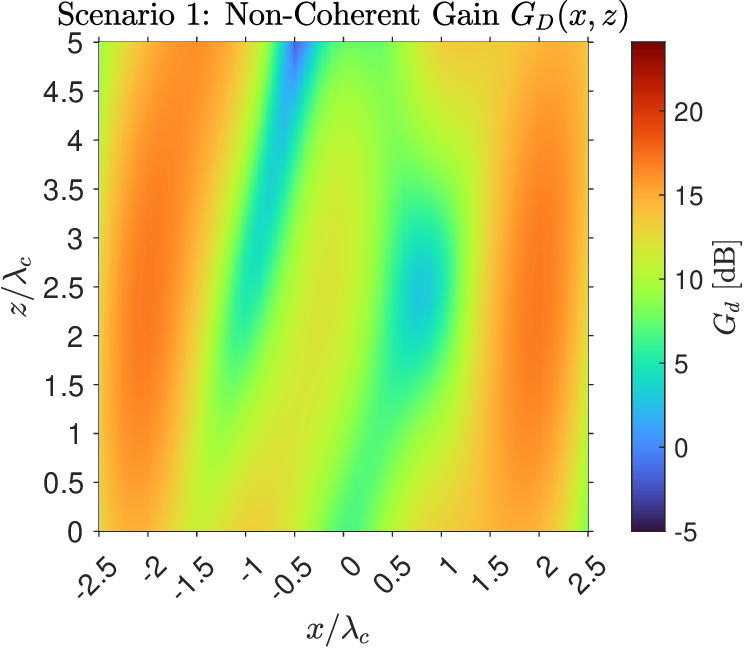}
		\label{fig:scen_1_C_0_XY_0}
	}
	\caption{Non-coherent combining gain distribution, $\gamma_m =1$ of  in different planes}
	\label{fig:scen_1_NC}
	\vspace{-10pt}
\end{figure}

The analogue weights $w_{m,n_x,n_y}$ are set to steer the array beams towards the user location $\ma{p}=\ma{0}$ using \ac{MRT}, while the digital precoding coefficients $\gamma_m$ are selected to achieve coherent combining at $\ma{p}$ under ideal synchronization.
A displacement from the intended focal point introduces relative phase shifts across the satellites, which reduces the coherent combining gain. To illustrate this effect, the gain distribution is evaluated around the intended focal point using a spatial sampling step of $\lambda_c/8$.
\figref{fig:scen_1_c} illustrates the gain distribution in a small region  around the intended focus point. 
It can be observed that the coherent combining region, which is defined as the area where the coherent combining gain remains close to the maximum value,  is confined approximately within $\pm \lambda_c/4$ along the $x$-axis, corresponding to a relative phase variation of approximately $\pm \pi/4$, consistent with the analysis in \figref{fig:gain_distribution_M_16}.
Along the $z$-axis, the gain variation is less sensitive, and the coherent region extends over several wavelengths. Outside the small focus region, the gain becomes similar to non-coherent combining, as illustrated in \figref{fig:scen_1_NC}, where the digital gains are  set to the uniform value $\gamma_m=1$
In this case, the gain distribution is governed by the uncontrolled phase differences between the received signals from different satellites.
In the case of the circular constellation, 
the coherent combining region is approximately confined within a radius of $\lambda_c/4$ in the $XY$ plane while extending over several wavelengths along the  $z$-axis. Since the user is located at the center of the circular geometry, the propagation delays from all satellites are nearly identical, thereby reducing the need for delay compensation across satellites.

\begin{figure}[t]
	\vspace{0.05in}
	\centering
	
	\subfloat[Gain in the $XY$ plane]{
		\includegraphics[width=0.23\textwidth]{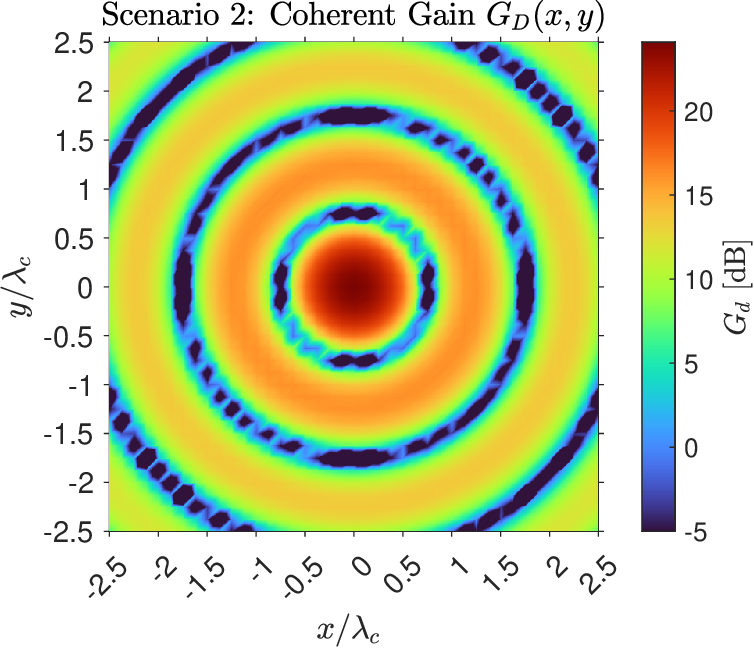}
		\label{fig:scen_2_C_1_XY_1}
	}
	\subfloat[Gain in the $YZ$ plane]{
		\includegraphics[width=0.23\textwidth]{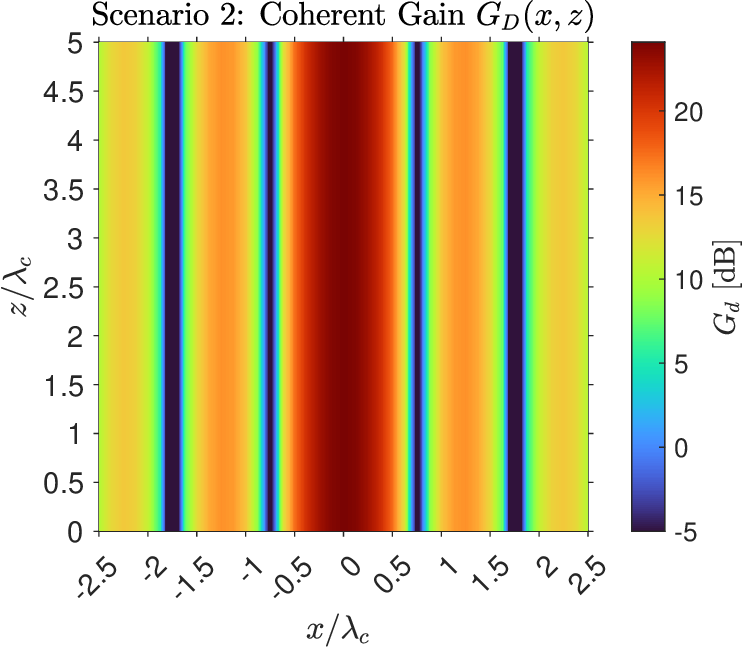}
		\label{fig:scen_2_C_1_XY_0}
	}
	\caption{Coherent combining gain of circular constellation at $\ma{p}=0$ and the resulting gain distribution in different planes}
	\label{fig:scen_2}
	\vspace{-10pt}
\end{figure}

\subsection{Potential of multi-user}
\begin{figure}[t]
	\vspace{0.05in}
	\centering	
	\subfloat[Linear constellation in the $XY$ plane]{
		\includegraphics[width=0.24\textwidth]{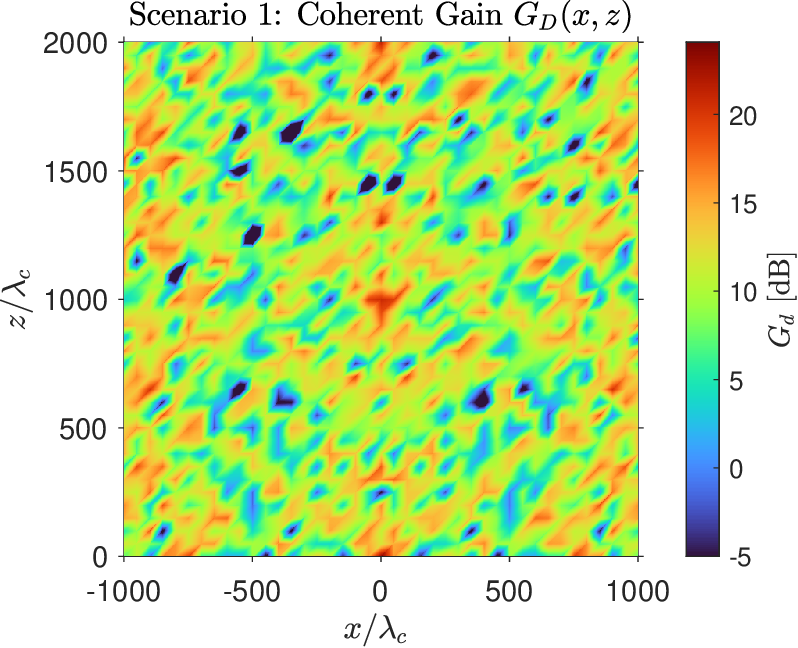}
		\label{fig:Large_area_scen_1_C_1_XY_0}
	}
	\subfloat[Circular constellation in the $YZ$ plane]{
		\includegraphics[width=0.24\textwidth]{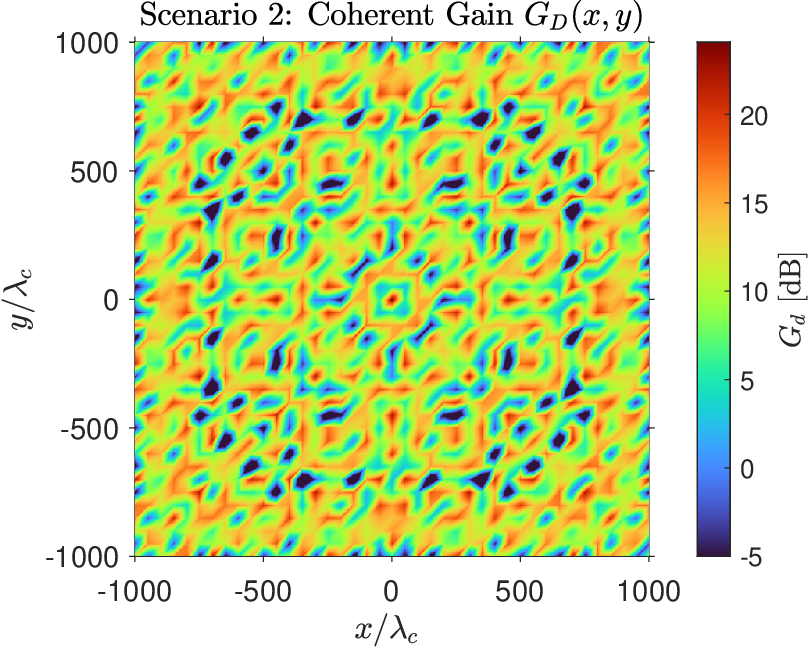}
		\label{fig:Large_area_scen_2_C_1_XY_1}
	}
	\caption{Coherent combining gain distribution over a larger area}
	\label{fig:large_scen_12}
	\vspace{-10pt}
\end{figure}

The linear constellation provides spatial separation mainly along the $XZ$ plane, enabling multi-user focusing in the elevation dimension.
In contrast, the circular constellation is more suitable for user separation over the $XY$ plane due to its azimuth diversity.
However, the sidelobes associated with \ac{MRT}-based focusing remain significant, as illustrated in \figref{fig:large_scen_12}. Thus, even if the arrays are divided into multiple subarrays to steer toward different locations, the resulting interference remains significant over many positions. Therefore, a joint digital precoding and analogue beamforming approach is necessary to control the gain distribution over the area.
Such approaches trade part of the maximum coherent gain for improved spatial selectivity and reduced interference.
\section{Conclusion}\label{sec:conclusions}
This paper analyzed distributed beam focusing for \ac{NTN} systems considering both linear and circular satellite constellations. The impact of phase errors resulting from synchronization and localization inaccuracies was investigated. The analysis showed that \ac{MRT}-based focusing enables a quadratic coherent combining gain relative to a single satellite. However, achieving such gain requires accurate phase alignment, corresponding approximately to synchronization accuracy within $\pm \frac{1}{8f_c}$ and positioning accuracy within $\pm \frac{\lambda_c}{8}$.
The considered constellation geometries provide different focusing characteristics. The circular constellation enables spatial focusing over the ground plane, while the linear constellation is more suitable for focusing in the elevation dimension. However, \ac{MRT}-based focusing results in strong sidelobes, which limit the applicability of spatial division within the covered area, even when multiple beams are generated at each satellite.

Therefore, further investigations of joint analog beamforming and digital precoding optimization are needed to control the gain distribution over the coverage area. Such approaches can relax the strict phase alignment requirements by enlarging the coherent focusing region, while simultaneously improving interference management and enabling more efficient spatial division within the constellation coverage area.

	\bibliographystyle{IEEEtran}
	\bibliography{references}

	\end{document}